\documentclass[showpacs,prb,twocolumn]{revtex4}%
\usepackage{amsfonts}
\usepackage{amsmath}
\usepackage{amssymb}
\usepackage{graphicx}
\usepackage{times}%
\providecommand{\U}[1]{\protect\rule{.1in}{.1in}}

\begin{document}
\title{Nonuniform Rashba-Dresselhaus spin precession along arbitrary paths}
\author{Ming-Hao Liu}
\affiliation{Department of Physics, National Taiwan University, Taipei 10617, Taiwan}
\author{Ching-Ray Chang}
\affiliation{Department of Physics, National Taiwan University, Taipei 10617, Taiwan}

\pacs{72.25.Dc, 71.70.Ej, 85.75.Hh}

\begin{abstract}
Electron spin precession in nonuniform Rashba-Dresselhaus two-dimensional
electron systems along arbitrary continuous paths is investigated. We derive
an analytical formula to describe the spin vectors (expectation values of the
injected spin) in such conditions using a contour-integral method. The
obtained formalism is capable of dealing with the nonuniformity of the Rashba
spin-orbit field due to the inherent random distribution of the ionized
dopants, and can be applied to curved one-dimensional quantum wires.
Interesting examples are given, and the modification to the spin precession
pattern in a Rashba-Dresselhaus channel when taking the random Rashba field
into account is shown.

\end{abstract}
\date{\today}
\maketitle

\section{Introduction}

Space-inversion-asymmetry-induced spin precession in two-dimensional electron
gas (2DEG) channels opens the possibility of spin-field-effect transistor
(SFET) (Ref. \onlinecite{Datta-Das}) and is itself an intriguing physical
phenomenon in the emerging field of semiconductor spintronics.\cite{SM
spintronics,RMP: spintronics} The heart of such spin precession lies in the
well-known Kramers theorem,\cite{Kittel} which predicts an intrinsic
zero-field spin-splitting in inversion-asymmetric crystal structures in the
absence of applied magnetic fields. The corresponding spin-splitting can be
described by an effective magnetic field, forcing the transport spins to
precess about such fields.

Mechanisms of inversion asymmetry in 2DEG systems include structure inversion
asymmetry, more commonly referred to as the Rashba spin-orbit (RSO)
interaction,\cite{Rashba term} of the confining potential, and the bulk
inversion asymmetry, referred to as the Dresselhaus spin-orbit (DSO) term for
zinc-blende semiconductors,\cite{Dresselhaus term} of the underlying crystal
structure. Specifically, the Rashba term originates from the transformation of
the electric field felt by the electron to an effective momentum-dependent
magnetic field. Hence factors influencing the RSO strength involve local
charge distribution (such as charged impurities\cite{Mel'nikov and Rashba} or
ionized dopant atoms\cite{Sherman sym qw,Sherman asym qw}), and applied
external electric fields (usually the gate voltage).

Perhaps the gate-voltage tunability of the RSO strength, quite successfully
demonstrated in InGaAs/InAlAs-based (Ref. \onlinecite{VGtunibility InGaAs})
and GaAs/AlAs-based (Refs. \onlinecite{VG aSO mstar} and
\onlinecite{VGtunibility GaAs}) quantum wells (QWs), determines the precession
rate of the injected electron spin, and is therefore the most important key to
the realization of the SFET. However, more than a decade of struggle has still
yielded disappointing results. Its failure may be ascribed to the so far less
successful factors:\cite{RMP: spintronics} (i) efficient spin injection rate
and (ii) uniformity of the RSO interaction. Apart from the most challenging
issue---the spin injection problem, nonuniformity of the RSO coupling arising
from the inherent random distribution of ionized dopants in the vicinity of
the 2DEG plane has recently attracted certain attention to reexamine the
Rashba 2DEG systems.\cite{Sherman sym qw,Sherman asym qw} This implies that
the RSO coupling strength $\alpha$ should be treated as position dependent,
and the corresponding modification to, e.g., spin precession rate, in the
Rashba channel must be taken into account.

In this paper we construct a contour-integral method to deal with spin
precession in nonuniform Rashba-Dresselhaus (RD) 2DEGs along arbitrary paths.
We derive an analytical formula to describe the spin vectors, i.e., the
expectation values of spin operators done with respect to the injected
electron spin state, inside the 2DEG channel. The obtained formula is found to
be a direct generalization of the previous one,\cite{MHL} which is applicable
to injected spins with straight space evolution in uniform RD 2DEGs. The
constructed formalism can straightforwardly deal with the nonuniformity of the
RSO coupling, DSO coupling, and even the effective mass $m^{\star}$, and is
applicable to curved one-dimensional (1D) channels, e.g., quantum ring
structures. Throughout this paper, single-particle quantum mechanics is used
and the effective mass approximation is adopted.

\section{Formula}

We begin by considering a [001]-grown zinc-blende-based 2DEG, where the Rashba
and Dresselhaus Hamiltonians are described by $H_{R}=\alpha/\hbar\left(
p_{y}\sigma^{x}-p_{x}\sigma^{y}\right)  $ and $H_{D}=\beta/\hbar\left(
p_{x}\sigma^{x}-p_{y}\sigma^{y}\right)  $, respectively. Whereas the RSO
strength $\alpha$ stems from the electric field generated by the ionized
dopants in the vicinity of the 2DEG plane, $\alpha$ is in principle position
dependent and material specific. On the other hand, the DSO strength, in bulk
systems, arises from the inversion-asymmetry of the crystal structure and thus
may not vary with different positions if the sample is well grown. However,
when restricted to 2DEGs, the DSO parameter $\beta$ depends on the width of
the confining quantum well, and may therefore also be position dependent. When
the underlying crystal structure of the 2DEG channel is not perfectly grown,
spatially varying effective mass $m^{\star}\left(  \vec{r}\right)  $ may also occur.

In the uniform case, where all of the three parameters $\alpha$, $\beta$, and
$m^{\star}$ are constant in space, an electron spin injected at $\vec{r}_{i}$
can be described by (see Appendix A)%
\begin{align}
\left\vert s\right\rangle _{\vec{r}_{i}\rightarrow\vec{r}_{d}}  & =e^{i\bar
{k}\left\vert \vec{r}_{d}-\vec{r}_{i}\right\vert }\sum_{\sigma=\pm}%
e^{-i\sigma\Delta\theta_{\vec{r}_{d}-\vec{r}_{i}}/2}\nonumber\\
& \times c_{\sigma}\left\vert \psi_{\sigma};\phi_{\vec{r}_{d}-\vec{r}_{i}%
}\right\rangle \left\langle \psi_{\sigma};\phi_{\vec{r}_{d}-\vec{r}_{i}%
}|s\right\rangle _{\vec{r}_{i}}\text{,}\label{sket_uniform}%
\end{align}
when detected at $\vec{r}_{d}$ after a straight space evolution $\vec{r}%
_{i}\rightarrow\vec{r}_{d}$. Here, $\left\vert s\right\rangle _{\vec{r}_{i}}$
is a given spinor describing the spin injected at $\vec{r}_{i}$, $\bar{k}$ is
the average of the two projected spin-dependent wave vector given by Eq.
(\ref{kbar}), and%
\begin{equation}
\left\vert \psi_{\sigma};\phi_{\vec{r}_{d}-\vec{r}_{i}}\right\rangle
\doteq\frac{1}{\sqrt{2}}\left(
\begin{array}
[c]{c}%
ie^{-i\varphi\left(  \alpha,\beta,\phi_{\vec{r}_{d}-\vec{r}_{i}}\right)  }\\
\sigma
\end{array}
\right)  \label{RDeigenspinor}%
\end{equation}
is the RD eigenspinor with $\phi_{\vec{r}_{d}-\vec{r}_{i}}$ being the argument
of the displacement vector $\vec{r}_{d}-\vec{r}_{i}$ and $\varphi\left(
\alpha,\beta,\phi\right)  \equiv\arg\left[  \left(  \alpha\cos\phi+\beta
\sin\phi\right)  +i\left(  \alpha\sin\phi+\beta\cos\phi\right)  \right]  $.
The phase difference is defined by
\begin{equation}
\Delta\theta_{\vec{r}}=\frac{2}{\hbar^{2}}m^{\star}\zeta\left(  \phi_{\vec{r}%
}\right)  r\text{ ,}\label{phase diff}%
\end{equation}
where
\begin{equation}
\zeta\left(  \phi_{\vec{r}}\right)  \equiv\sqrt{\alpha^{2}+\beta^{2}%
+2\alpha\beta\sin\left(  2\phi_{\vec{r}}\right)  }\label{so strength}%
\end{equation}
is the composite spin-orbit coupling strength. Note that in this paper the
spatially evolved state ket without superscript, say $\left\vert
s\right\rangle _{\vec{r}_{1}\rightarrow\vec{r}_{2}}$, means space evolution
from position $\vec{r}_{1}$ to position $\vec{r}_{2}$ through the straight
path connecting the two points, while $\left\vert s\right\rangle _{\vec{r}%
_{1}\rightarrow\vec{r}_{2}}^{C}$ means evolution through path $C$.%
\begin{figure}
[ptb]
\begin{center}
\includegraphics[
height=1.5791in,
width=3.3027in
]%
{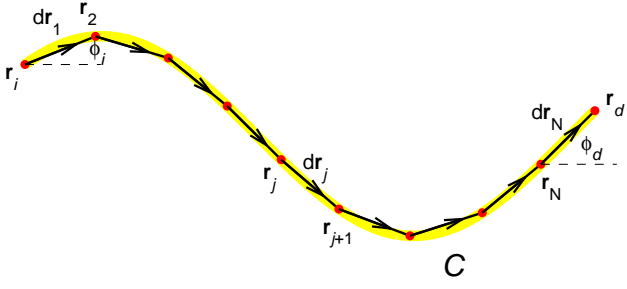}%
\caption{(Color online) Schematic of a continuous curved path sectioned into
$N$ pieces.}%
\label{FIG_path}%
\end{center}
\end{figure}

For the nonuniform case, we consider an arbitrary continuous path $C$
connecting the injection and detection points, and section the path into $N$
pieces (see Fig. \ref{FIG_path}). In the limit of $N\rightarrow\infty$, each
section approaches to a straight path with constant $\alpha$, $\beta$, and
$m^{\star}$, and Eqs. (\ref{sket_uniform}) and (\ref{phase diff}) are then
applicable for each section. Again, we inject a spin at $\vec{r}_{i}$ and
detect the spin at $\vec{r}_{d}$ through, however, the curved path $C$ along
which the parameters $\alpha$, $\beta$, and $m^{\star}$ may be functions of
position. After successive application of Eq. (\ref{sket_uniform}), we obtain
(see Appendix B)%
\begin{equation}
\left\vert s\right\rangle _{\vec{r}_{i}\rightarrow\vec{r}_{d}}^{C}%
=\sum_{\sigma}\exp\left(  -\frac{i\sigma}{2}\Delta\Theta\right)  \left\langle
\psi_{\sigma};\phi_{i}|s\right\rangle _{\vec{r}_{i}}\left\vert \psi_{\sigma
};\phi_{d}\right\rangle \text{,} \label{sket_nonuni}%
\end{equation}
where the total phase difference is given by a contour integral%
\begin{equation}
\Delta\Theta=\frac{2}{\hbar^{2}}\int_{C}m^{\star}\zeta d\vec{r}\text{ .}
\label{total phase diff}%
\end{equation}
Compared to the uniform case Eqs. (\ref{sket_uniform}) and (\ref{phase diff}),
the problem of nonuniform systems is simply extended to integrate the total
phase difference encountered by the electron along the path $C$.

We remind here that the above formalism (and later the derived spin vector
formula) is applicable for nonuniform systems but only with continuous spatial
evolution. Propagation along discontinuous paths must be handled by sectioning
the path into pieces of continuous ones [since Eq. (\ref{spinor overlap}) does
not hold on discontinuous points]. Another discontinuity that may crash the
above formalism is the reverse of the RD field direction [at which Eq.
(\ref{spinor overlap}) also fails]. In pure Rashba (such as Si-Si$_{x}%
$Ge$_{1-x}$ asymmetric QWs) cases, the change of sign $\alpha\rightarrow
-\alpha$, due to either the random dopants or the reverse gating induces such
discontinuity. In the composite case (both $\alpha$ and $\beta$ nonvanishing),
the cancellation of the momentum dependence of the RD effective magnetic field
at the condition $\left\vert \alpha\right\vert \approx\left\vert
\beta\right\vert $ provides possibility to run the SFET in the nonballistic
regime \cite{NBSFET} and also contains fundamental physical
phenomena.\cite{sign change} However, the corresponding field direction may
become unstable along $\pm$[1\={1}0] or $\pm$[110] directions, even though the
field magnitude is still continuous. When dealing with such systems, one must
carefully check the continuity of the RD field direction and do the
calculation partitively.

Now we apply Eq. (\ref{sket_nonuni}) to perform the spin expectation values
$\langle\vec{S}\rangle\equiv\hbar/2\left\langle \vec{\sigma}\right\rangle $,
where $\vec{\sigma}$ is the Pauli matrix vector $\left(  \sigma_{x},\sigma
_{y},\sigma_{z}\right)  $. After some mathematical manipulation, we
obtain\begin{widetext}%
\begin{equation}
\left\langle \vec{\sigma}\right\rangle =\left(
\begin{array}
[c]{c}%
-\cos\theta_{s}\cos\varphi_{d}\sin\Delta\Theta+\sin\theta_{s}\left[
\cos\left(  \varphi_{d}-\varphi_{i}+\phi_{s}\right)  \cos^{2}\dfrac
{\Delta\Theta}{2}-\cos\left(  \varphi_{d}+\varphi_{i}-\phi_{s}\right)
\sin^{2}\dfrac{\Delta\Theta}{2}\right]  \\
-\cos\theta_{s}\sin\varphi_{d}\sin\Delta\Theta+\sin\theta_{s}\left[
\sin\left(  \varphi_{d}-\varphi_{i}+\phi_{s}\right)  \cos^{2}\dfrac
{\Delta\Theta}{2}-\sin\left(  \varphi_{d}+\varphi_{i}-\phi_{s}\right)
\sin^{2}\dfrac{\Delta\Theta}{2}\right]  \\
\cos\theta_{s}\cos\Delta\Theta+\sin\theta_{s}\cos\left(  \varphi_{i}-\phi
_{s}\right)  \sin\Delta\Theta
\end{array}
\right)  \text{,}\label{<S>}%
\end{equation}
\end{widetext}with $\varphi_{i\left(  d\right)  }\equiv\arg[(\alpha_{i\left(
d\right)  }\cos\phi_{i\left(  d\right)  }+\beta_{i\left(  d\right)  }\sin
\phi_{i\left(  d\right)  })+i(\alpha_{i\left(  d\right)  }\sin\phi_{i\left(
d\right)  }+\beta_{i\left(  d\right)  }\cos\phi_{i\left(  d\right)  })]$,
$\theta_{s}$ and $\phi_{s}$ being the polar and azimuthal angles of the
injected spin, respectively. Clearly, Eq. (\ref{<S>}) recovers the previous
spin vector formula \cite{MHL} when $\alpha$, $\beta$, $m^{\star}$, and $\phi$
are constants, i.e., uniform spin precession along straight paths.

\section{Results and Discussion}

In this section we apply our generalized version of spin-vector formula Eq.
(\ref{<S>}) with the total phase $\Delta\Theta$ given by Eq.
(\ref{total phase diff}) first to some interesting 1D geometries, including
quantum rings and quantum wires, and then to a more realistic 2DEG case, where
position dependent RSO coupling, generated by regular and random distributions
of ionized dopants, is present. In all cases, we assume constant DSO coupling
and the electron effective mass.

\subsection{1D quantum ring}

Beginning with an ideal Rashba half-ring with ring radius $0.1$ $%
\operatorname{\mu m}%
$, we inject an electron at one end, and detect $\langle\vec{S}\rangle$ down
the way along the ring to the other end. The half-ring is assumed to be
ideally 1D and made of InGaAs-based materials. We set the RSO parameter
$\alpha=0.03%
\operatorname{eV}%
\operatorname{nm}%
$ and the electron effective mass $m^{\star}=0.03m_{e}$.\cite{Sato}%
\begin{figure}
[pb]
\begin{center}
\includegraphics[
height=3.6746in,
width=3.0761in
]%
{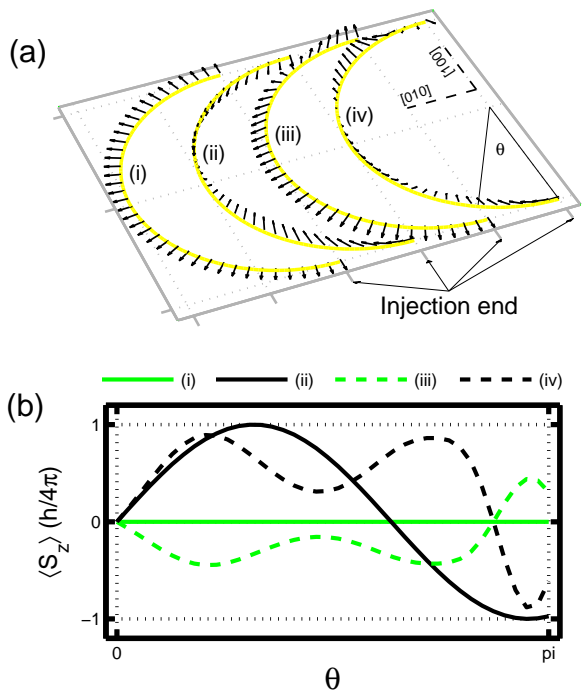}%
\caption{(Color online) (a) Spin vectors along four ideal 1D half-rings. Rings
(i) and (ii) contains only the RSO while in rings (iii) and (iv) the DSO is
involved. The spin, polarized parallel either to the radial or the tangential
directions, is injected at the lower end of each ring. (b) Corresponding
$\left\langle S_{z}\right\rangle $ in units of $\hbar/2$ as a function of the
ring argument $\theta$.}%
\label{FIG_QR}%
\end{center}
\end{figure}
Two configurations of the injected spin are considered. In ring (i) of Fig.
\ref{FIG_QR}(a), we inject the spin with polarization parallel to the radial
direction, which is also the eigenstate of this Rashba ring case. As expected,
the spin vectors maintain in the radial direction since the spin is injected
in its eigenstate and will stay in this state hereafter. This can also be
viewed as a generalized version of our previous proposal of precessionless
spin transport wire.\cite{MHLjap} In Rashba systems, one can precessionlessly
transport spins even through an arbitrary wire shape (but with continuous
curvature), once the spin is injected with polarization perpendicular to the
wire direction.

Next we inject the spin parallel to the tangential direction. As shown in ring
(ii) of Fig. \ref{FIG_QR}(a), the injected spin precesses upright down the way
to the end. This is equivalent to the Datta-Das SFET with a curved 1D Rashba
channel. Whereas the RSO coupling induces an effective magnetic field $\vec
{B}_{\text{eff}}$, which is always perpendicular to the electron transport
($\vec{B}_{\text{eff}}$ $\parallel$ the radial direction here), rings (i) and
(ii) in Fig. \ref{FIG_QR}(a) are both reasonable and expectable. However, when
the DSO coupling is involved, the $z$-rotational symmetry is broken, and the
spin vectors cannot be expected intuitively, but can still be well described
by our formula Eq. (\ref{<S>}) [see rings (iii) and (iv) in Fig.
\ref{FIG_QR}(a)]. To clarify the influence due to the DSO, we plot
$\left\langle S_{z}\right\rangle $ as a function of the ring argument $\theta$
in Fig. \ref{FIG_QR}(b), where line (i) and line (ii) exhibit precessionless
and complete precessing behaviors, respectively, and change to line (iii) and
line (iv) after getting the DSO coupling involved.%
\begin{figure}
[ptb]
\begin{center}
\includegraphics[
height=1.8844in,
width=3.3477in
]%
{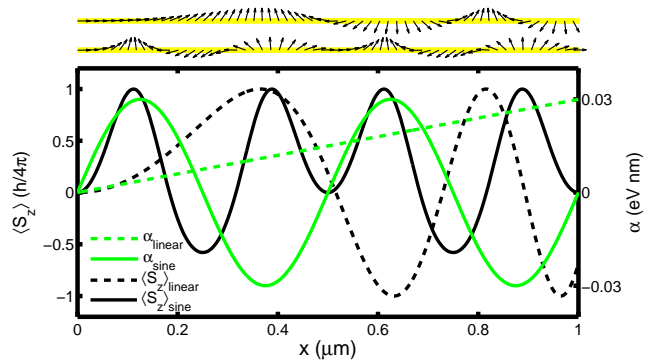}%
\caption{(Color online) Two ideal 1D wires with position-dependent RSO
strength. The upper wire shows the effect of a linear $\alpha$ and the lower
wire shows a sine-varying $\alpha$. In the main panel, $\alpha$ and
$\left\langle S_{z}\right\rangle $ as functions of the longitudinal position
are plotted.}%
\label{FIG_QW}%
\end{center}
\end{figure}

\subsection{1D quantum wire}

We now consider two straight quantum wires with position-dependent RSO
strength, one being a linear $\alpha$ wire and the other a sine-varying
$\alpha$ wire. These wires exhibit interesting spin precession behaviors, as
shown in Fig. \ref{FIG_QW}, where the upper and lower wires, sketched from the
side view, correspond to the linear and sine-varying Rashba wires,
respectively. As expected, the injected spin rotates with a precession rate
increasing with the position, in the linear Rashba wire. In the sine case, the
alternate positive and negative RSO drives the spin to precess backward (when
$\alpha>0$) and forward (when $\alpha<0$). These quantum wires may be
fabricated by special gating designs, but applicability to the spintronic
device is not obvious. However, the physics indicated here is simple and
clear. Whereas the magnitude of the RSO strength $\alpha$ determines the spin
precession rate, the sign of $\alpha$ controls the precession direction. From
the viewpoint of gating, a strong enough backgate may reverse the electric
field direction in the 2DEG, and hence the resulting Rashba field direction,
forcing the spin to precess in an opposite direction. From the viewpoint of
the random Rashba field, such an alternate $\alpha$ is similar to that in a
symmetric quantum well, e.g., a Si-Si$_{x}$Ge$_{1-x}$ well, and will induce a
finite spin relaxation rate.\cite{Sherman sym qw}

\subsection{2DEG channel with regular and random dopants: Random Rashba
effect}

Finally, we arrive at the crucial but realistic problem: how will the spin
precession pattern be modified due to the random Rashba field? A first guess
on the basis of Eq. (\ref{total phase diff}) is that how wide the angle the
spin precesses through lies in how much spin-orbit interaction the electron
encounters along the path it goes. A symmetric well, which has a vanishing
spatially averaged RSO $\left\langle \alpha\right\rangle =0$, does not have
spin precession pattern within a submicrometer scale, even though there are
nonvanishing fluctuating Rashba fields. This is similar to the sine-varying
Rashba wire shown in Fig. \ref{FIG_QW} but with a much rapider $\alpha$
oscillation period ($\sim$ atomic sizes). We now consider an asymmetric well,
and adopt the same principle introduced by Sherman \textit{et al.}
\cite{Sherman asym qw} to generate the random Rashba field.%
\begin{figure}
[ptb]
\begin{center}
\includegraphics[
height=3.3425in,
width=3.4566in
]%
{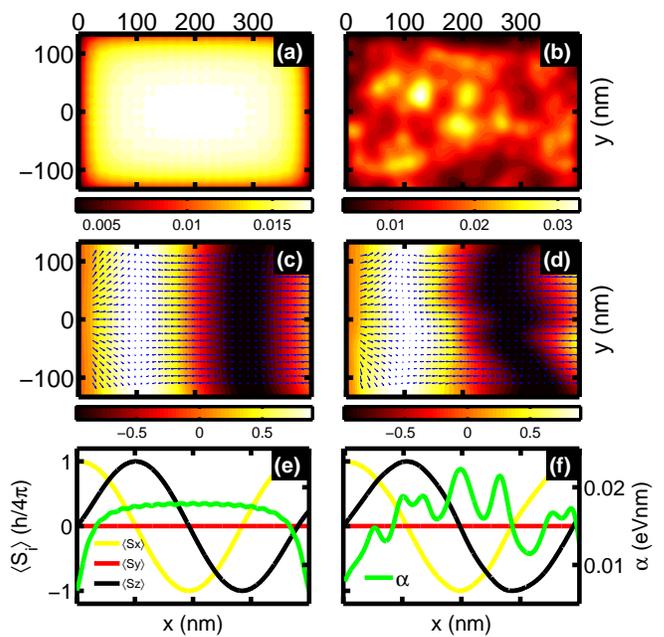}%
\caption{(Color online) Position-dependent Rashba field $\alpha$ with (a)
regular and (b) random dopant distribution, resulting in the corresponding
spin precession patterns in (c) and (d), respectively, where point injection
of $x$-polarized spins at the left ends, are assumed. The color bars determine
the magnitude of $\alpha$ in units of $\operatorname{eV}\operatorname{nm}$ in
(a) and (b), and the $z$ compont of the spin vector $\left\langle
S_{z}\right\rangle $ in units of $\hbar/2$ in (c) and (d). (e) and (f) show
$\left\langle S_{x}\right\rangle ,$ $\left\langle S_{y}\right\rangle ,$
$\left\langle S_{z}\right\rangle ,$ and $\alpha$ as functions of longitudinal
position $x$ at $y=0$ for the regular and the random cases, respectively.}%
\label{FIG_randomRashba}%
\end{center}
\end{figure}

Specifically, we consider a $400%
\operatorname{nm}%
\times267%
\operatorname{nm}%
$ InAs-based 2DEG with only a single dopant layer, located at a spacing
$z_{0}=$ $20$ $%
\operatorname{nm}%
$ from the conducting plane. Within the dopant layer, charged dopant atoms are
randomly distributed, generating the Coulomb field with $z$ component
described by $E_{z}\left(  \vec{\rho}\right)  =ez_{0}/\epsilon\sum_{j}\left[
\left(  \vec{\rho}-\vec{r}_{j}\right)  ^{2}+z_{0}^{2}\right]  ^{-3/2}$. Here
$e$ is the electron charge, $\epsilon$ is the material permittivity,
$\vec{\rho}\equiv\left(  x,y\right)  $ is the two-dimensional position vector,
and $\vec{r}_{j}$ is the position of the $j$th dopant. We take the dopant
concentration $\bar{n}=2.5\times10^{11}$ $%
\operatorname{cm}%
^{-2}$ (same with Ref. \onlinecite{Sherman asym qw}), the permittivity
$\epsilon=15.1\epsilon_{0}$ ($\epsilon_{0}$ is the free space permittivity),
and the lattice constant $a=0.6058$ $%
\operatorname{nm}%
$.\cite{book} Using the linear model $\alpha\left(  \vec{\rho}\right)
=\alpha_{\text{SO}}eE_{z}\left(  \vec{\rho}\right)  $ with $\alpha_{\text{SO}%
}=110$ $%
\operatorname{\text{\AA}}%
^{2}$ for InAs,\cite{VG aSO mstar,aSO} we obtain the Rashba field in the cases
of regular and random dopant distributions in Figs. \ref{FIG_randomRashba}(a)
and \ref{FIG_randomRashba}(b), respectively.

In the 2DEG channel, we also take into account the DSO coupling by assuming
the well thickness $w=50$ $%
\operatorname{\text{\AA}}%
$, leading to $\beta\approx\gamma\left\langle k_{z}^{2}\right\rangle
\approx1.0\allowbreak62\times10^{-3}$ $%
\operatorname{eV}%
\operatorname{nm}%
$. Here $\gamma=26.9$ $%
\operatorname{eV}%
\operatorname{\text{\AA}}%
^{3}$ for InAs QWs (Ref. \onlinecite{VG aSO mstar}) and $\left\langle
k_{z}^{2}\right\rangle \approx\left(  \pi/w\right)  ^{2}$ in the case of rigid
walls. Assuming point injection of $x$-polarized spins at the left-center of
the channel with transport direction [110], the spin vector formula Eq.
(\ref{<S>}) gives the spin precession patterns, as shown in Fig.
\ref{FIG_randomRashba}(c) for the regular channel and Fig.
\ref{FIG_randomRashba}(d) for the random channel. Clearly, the randomness of
the dopant distribution distorts the precession pattern, and hence is expected
to lower the purity of the spin signal collected at the drain end. This may
bring another difficulty in realizing the Datta-Das
transistor.\cite{Datta-Das} Moreover, such unpredictable random distribution
will eventually cause an uncertainty of the collected spin signal, for each
individual 2DEG channel, even if the dopant concentration is perfectly
controlled. For example, the standard deviation calculation with the 2DEG
channel conditions considered here shows that $\Delta S_{x}/\left(
\hbar/2\right)  \sim$ $\mathcal{O}\left(  10^{-1}\right)  $ for InAs-based and
$\Delta S_{x}/\left(  \hbar/2\right)  \sim\mathcal{O}\left(  10^{-2}\right)  $
for GaAs-based materials. (Here $\Delta S_{i}$ means the standard deviation of
the averaged $i$ component of the spin signal collected at the drain end in
the channel, among a number of different samplings.)

Note that we have chosen [110] as the channel direction, where the Rashba and
Dresselhaus fields are parallel,\cite{MHL} and the random effect seems
moderate. When analyzing other channel directions such as [1\={1}0], the
obtained spin precession pattern will be totally different from the regular
case. Last, if we go back to the 1D case and focus on the center path from the
source end to the drain end, the distortion of the spin precession curves due
to the random Rashba field is surprisingly weak [see Figs.
\ref{FIG_randomRashba}(e) and \ref{FIG_randomRashba}(f)]. This again supports
the suggestion of using 1D or quasi-1D channels for the Datta-Das
transistor.\cite{Datta-Das,Confinement}

\section{Summary}

In conclusion, we have derived an analytical formula to describe the spin
precession in nonuniform RD 2DEGs, using a contour-integral method. The
obtained results are applicable for 2DEG systems with position-dependent
$\alpha$, $\beta$, and $m^{\star}$, and curved 1D wires. Numerical examples
show interesting spin precession behaviors, and also support the idea of the
precessionless spin transport wire\cite{MHLjap} in the Rashba ring case. We
have also demonstrated how the spin precession pattern in a RD 2DEG may be
distorted when taking the random Rashba field into account. In the viewpoint
of carrying out the SFET, the random Rashba effect may distort the regular
spin precession pattern in the 2DEG channel, especially along weak spin-orbit
strength directions. Such undesired influence can be suppressed quite well
when narrowing down the channel to 1D [compare Figs. \ref{FIG_randomRashba}(f)
with \ref{FIG_randomRashba}(e)], supporting the pioneering suggestions of
using quasi-1D channels.\cite{Datta-Das,Confinement}

\begin{acknowledgments}
One of the authors (M.H.L.) is grateful to Son-Hsien Chen and Ivo Klik for
valuable discussions and suggestions. This work is supported by the Republic
of China National Science Council Grant No. 94-2112-M-002-004.
\end{acknowledgments}

\appendix{}

\section{Uniform space evolution along straight path}

To derive Eq. (\ref{sket_uniform}), we use the translation operator to bring
the ket $\left\vert s\right\rangle _{\vec{r}_{i}}$ from $\vec{r}_{i}$ to
$\vec{r}_{d}$: $\left\vert s\right\rangle _{\vec{r}_{i}\rightarrow\vec{r}_{d}%
}=e^{i\vec{p}\cdot\left(  \vec{r}_{d}-\vec{r}_{i}\right)  /\hbar}\left\vert
s\right\rangle _{\vec{r}_{i}}$. Expanding $\left\vert s\right\rangle _{\vec
{r}_{i}}$ in terms of the RD eigenspinors $\left\vert \psi_{\pm};\phi_{\vec
{r}_{d}-\vec{r}_{i}}\right\rangle $ and denoting $\Delta r=\left\vert \vec
{r}_{d}-\vec{r}_{i}\right\vert $, we have%
\begin{equation}
\left\vert s\right\rangle _{\vec{r}_{i}\rightarrow\vec{r}_{d}}=e^{ip\Delta
r/\hbar}\left(  c_{+}\left\vert \psi_{+};\phi_{\vec{r}_{d}-\vec{r}_{i}%
}\right\rangle +c_{-}\left\vert \psi_{-};\phi_{\vec{r}_{d}-\vec{r}_{i}%
}\right\rangle \right)  \text{ ,}\label{A1}%
\end{equation}
where $c_{\pm}=\left\langle \psi_{\pm};\phi_{\vec{r}_{d}-\vec{r}_{i}%
}|s\right\rangle _{\vec{r}_{i}}$ are the expansion coefficients ($\phi
_{\vec{r}}$ is the argument angle of the vector $\vec{r}$). Note that the
electron carrying momentum $\vec{p}$ is supposed to move in the direction
$\vec{r}_{d}-\vec{r}_{i}$, so the product $\vec{p}\cdot\left(  \vec{r}%
_{d}-\vec{r}_{i}\right)  $ is simplified to $p\left\vert \vec{r}_{d}-\vec
{r}_{i}\right\vert =p\Delta r$. Defining%
\begin{equation}
k_{-}-k_{+}\equiv\Delta k=\frac{2m^{\star}\zeta}{\hbar^{2}}\text{ ,}%
\end{equation}
where $\zeta$ is the composite spin-orbit strength given in Eq.
(\ref{so strength}), we rewrite Eq. (\ref{A1}) as%
\begin{align}
\left\vert s\right\rangle _{\vec{r}_{i}\rightarrow\vec{r}_{d}} &
=e^{ik_{+}\Delta r}c_{+}\left\vert \psi_{+}\right\rangle +e^{ik_{-}\Delta
r}c_{-}\left\vert \psi_{-}\right\rangle \nonumber\\
&  =e^{i\bar{k}\Delta r}\sum_{\sigma=\pm}e^{-i\sigma\Delta\theta/2}c_{\sigma
}\left\vert \psi_{\sigma};\phi_{\vec{r}_{d}-\vec{r}_{i}}\right\rangle \text{
,}\label{sket_uniform_original}%
\end{align}
with
\begin{equation}
\bar{k}\equiv\frac{k_{+}+k_{-}}{2}\label{kbar}%
\end{equation}
and $\Delta\theta\equiv\Delta k\Delta r$. Note that the common phase
$e^{i\bar{k}\Delta r}$ will vanish when calculating expectation values.
Whereas $\bar{k}$ depends on the energy the electron spin carries while
$\Delta k$, or the phase difference $\Delta\theta$, depends only on the
spin-orbit strength, the physics of $\left\vert s\right\rangle _{\vec{r}%
_{i}\rightarrow\vec{r}_{d}}$, such as the spatial spin precession, does not
depend on the injection energy.

\section{Nonuniform space evolution along curved path}

To derive Eqs. (\ref{sket_nonuni}) and (\ref{total phase diff}), we refer to
Fig. \ref{FIG_path} where we denote $\vec{r}_{i\left(  d\right)  }\equiv
\vec{r}_{1\left(  N+1\right)  }$ and successively apply Eq.
(\ref{sket_uniform}) from the last to the first section. Taking the $j$th
section for example, we have%
\begin{equation}
\left\vert s\right\rangle _{\vec{r}_{j}\rightarrow\vec{r}_{j+1}}=e^{i\bar
{k}_{j}\Delta r}\sum_{\sigma=\pm1}e^{-i\sigma\Delta\theta_{j}/2}\left\langle
\psi_{\sigma};\phi_{j}|s\right\rangle _{\vec{r}_{j}}\left\vert \psi_{\sigma
};\phi_{j}\right\rangle \text{ ,}%
\end{equation}
with $\Delta r=\left\vert \vec{r}_{j+1}-\vec{r}_{j}\right\vert $, $\phi
_{j}\equiv\phi_{\vec{r}_{j+1}-\vec{r}_{j}}$, $\Delta\theta_{j}=2m_{j}^{\star
}\zeta_{j}\Delta r/\hbar^{2}$, and $\zeta_{j}\equiv\sqrt{\alpha_{j}^{2}%
+\beta_{j}^{2}+2\alpha_{j}\beta_{j}\sin\left(  2\phi_{j}\right)  }$. Note that
the RSO strength, DSO strength, and effective mass are ideally local:
$\alpha_{j}\equiv\alpha\left(  \vec{r}_{j}\right)  $, $\beta_{j}\equiv
\beta\left(  \vec{r}_{j}\right)  $, and $m_{j}^{\star}\equiv m^{\star}\left(
\vec{r}_{j}\right)  $. Note also that the common phase factor outside the
summation $\exp\left(  i\bar{k}_{j}\Delta r\right)  $ with $\bar{k}%
_{j}=\left(  k_{j}^{+}+k_{j}^{-}\right)  /2$ may depend not only on the
position $\vec{r}_{j}$ but also on the direction $\phi_{j}$ due to the
anisotropic Fermi contour in the general RD case. However, we will show that,
as in the uniform straight case, this phase does not contribute to the spin
vector $\langle\vec{S}\rangle$. In the following we will abbreviate the RD
eigenspinor $\left\vert \psi_{\sigma};\phi_{j}\right\rangle $ to $\left\vert
\psi_{\sigma}^{j}\right\rangle $ for simplicity.

For the full translation, we can start with the last section%
\begin{align}
\left\vert s\right\rangle _{\vec{r}_{i}\rightarrow\vec{r}_{f}}^{C} &
=\left\vert s\right\rangle _{\vec{r}_{N}\rightarrow\vec{r}_{N+1}}\nonumber\\
&  =e^{i\bar{k}_{N}\Delta r}\sum_{\sigma=\pm1}e^{-i\sigma\Delta\theta_{N}%
/2}\left\langle \psi_{\sigma}^{N}|s\right\rangle _{\vec{r}_{N-1}%
\rightarrow\vec{r}_{N}}\left\vert \psi_{\sigma}^{N}\right\rangle \text{
,}\label{B2}%
\end{align}
and successively substitute
\begin{subequations}
\label{0}%
\begin{align}
\left\vert s\right\rangle _{\vec{r}_{N-1}\rightarrow\vec{r}_{N}} &
=e^{i\bar{k}_{N-1}\Delta r}\sum_{\sigma=\pm1}e^{-i\sigma\Delta\theta_{N-1}%
/2}\nonumber\\
&  \times\left\langle \psi_{\sigma}^{N-1}|s\right\rangle _{\vec{r}%
_{N-2}\rightarrow\vec{r}_{N-1}}\left\vert \psi_{\sigma}^{N-1}\right\rangle \\
\left\vert s\right\rangle _{\vec{r}_{N-2}\rightarrow\vec{r}_{N-1}} &
=e^{i\bar{k}_{N-2}\Delta r}\sum_{\sigma=\pm1}e^{-i\sigma\Delta\theta_{N-2}%
/2}\nonumber\\
&  \times\left\langle \psi_{\sigma}^{N-2}|s\right\rangle _{\vec{r}%
_{N-3}\rightarrow\vec{r}_{N-2}}\left\vert \psi_{\sigma}^{N-2}\right\rangle \\
&  \vdots\nonumber\\
\left\vert s\right\rangle _{\vec{r}_{1}\rightarrow\vec{r}_{2}} &  =e^{i\bar
{k}_{1}\Delta r}\sum_{\sigma=\pm1}e^{-i\sigma\Delta\theta_{1}/2}\left\langle
\psi_{\sigma}^{1}|s\right\rangle _{\vec{r}_{1}}\left\vert \psi_{\sigma}%
^{1}\right\rangle
\end{align}
into the expression Eq. (\ref{B2}). Then totally we have%
\end{subequations}
\begin{align}
\left\vert s\right\rangle _{\vec{r}_{i}\rightarrow\vec{r}_{f}}^{C} &
=e^{i\bar{k}_{N}\Delta r}e^{i\bar{k}_{N-1}\Delta r}\cdots e^{i\bar{k}%
_{1}\Delta r}\sum_{\sigma^{\left(  N\right)  },\sigma^{\left(  N-1\right)
},\cdots\sigma^{\prime\prime},\sigma^{\prime}}\nonumber\\
&  \times e^{-i\sigma^{\left(  N\right)  }\Delta\theta_{N}/2}e^{-i\sigma
^{\left(  N-1\right)  }\Delta\theta_{N-1}/2}\cdots e^{-i\sigma^{\prime\prime
}\Delta\theta_{2}/2}\nonumber\\
&  \times e^{-i\sigma^{\prime}\Delta\theta_{1}/2}(\left\langle \psi
_{\sigma^{\left(  N\right)  }}\right\vert \left\langle \psi_{\sigma^{\left(
N-1\right)  }}\right\vert \cdots\left\langle \psi_{\sigma^{\prime\prime}%
}\right\vert \left\langle \psi_{\sigma^{\prime}}|s\right\rangle _{0}%
\nonumber\\
&  \times\left\vert \psi_{\sigma^{\prime}}\right\rangle \cdots\left\vert
\psi_{\sigma^{\left(  N-2\right)  }}\right\rangle \left\vert \psi
_{\sigma^{\left(  N-1\right)  }}\right\rangle )\left\vert \psi_{\sigma
^{\left(  N\right)  }}\right\rangle \text{ ,}\label{B4}%
\end{align}
where we have further abbreviated $\left\vert \psi_{\sigma^{\left(  j\right)
}}^{j}\right\rangle $ to $\left\vert \psi_{\sigma^{\left(  j\right)  }%
}\right\rangle $. One must bear in mind that the shorthand $\left\vert
\psi_{\sigma^{\left(  j\right)  }}\right\rangle $ actually means $\left\vert
\psi_{\sigma^{\left(  j\right)  }};\phi_{j}\right\rangle $. In Eq. (\ref{B4}),
the spinor overlaps can be rearranged as $\left\langle \psi_{\sigma^{\prime}%
}|s\right\rangle _{0}\left\langle \psi_{\sigma^{\prime\prime}}|\psi
_{\sigma^{\prime}}\right\rangle \cdots\left\langle \psi_{\sigma^{\left(
N-1\right)  }}|\psi_{\sigma^{\left(  N-2\right)  }}\right\rangle \left\langle
\psi_{\sigma^{\left(  N\right)  }}|\psi_{\sigma^{\left(  N-1\right)  }%
}\right\rangle $. For the $j$th bracket, the overlap reads%
\begin{equation}
\left\langle \psi_{\sigma^{\left(  j\right)  }}|\psi_{\sigma^{\left(
j-1\right)  }}\right\rangle =\tfrac{1}{2}\left(  e^{i\left(  \varphi
_{j}-\varphi_{j-1}\right)  }+\sigma^{\left(  j\right)  }\sigma^{\left(
j-1\right)  }\right)  \text{ .}%
\end{equation}
In the limit of $N\rightarrow\infty$ and the continuous condition, this
overlap becomes $\delta_{\sigma^{\left(  j\right)  },\sigma^{\left(
j-1\right)  }}$ since $\phi_{j}\approx\phi_{j-1},$ $\alpha_{j}\approx
\alpha_{j-1},$ and $\beta_{j}\approx\beta_{j-1}$ yield $\varphi_{j}%
\approx\varphi_{j-1}$, so that we have $\left\langle \psi_{\sigma^{\left(
j-1\right)  }}|\psi_{\sigma^{\left(  j\right)  }}\right\rangle \approx\left(
1+\sigma^{\left(  j\right)  }\sigma^{\left(  j-1\right)  }\right)
/2=\delta_{\sigma^{\left(  j\right)  },\sigma^{\left(  j-1\right)  }}$.
Therefore, the spinor overlaps are reduced to%
\begin{align}
&  \left\langle \psi_{\sigma^{\prime}}|s\right\rangle _{0}\left\langle
\psi_{\sigma^{\prime\prime}}|\psi_{\sigma^{\prime}}\right\rangle
\cdots\left\langle \psi_{\sigma^{\left(  N-1\right)  }}|\psi_{\sigma^{\left(
N-2\right)  }}\right\rangle \left\langle \psi_{\sigma^{\left(  N\right)  }%
}|\psi_{\sigma^{\left(  N-1\right)  }}\right\rangle \nonumber\\
&  =\delta_{\sigma^{\left(  N\right)  },\sigma^{\left(  N-1\right)  }}%
\delta_{\sigma^{\left(  N-1\right)  },\sigma^{\left(  N-2\right)  }}%
\cdots\delta_{\sigma^{\prime\prime},\sigma^{\prime}}\left\langle \psi
_{\sigma^{\prime}}|s\right\rangle _{0}\text{ ,}\label{spinor overlap}%
\end{align}
leading to a greatly simplified expression%
\begin{align}
\left\vert s\right\rangle _{\vec{r}_{i}\rightarrow\vec{r}_{d}}^{C}  &
=\exp\left(  i\sum_{n=1}^{N}\bar{k}_{n}\Delta r\right)  \nonumber\\
& \times\sum_{\sigma}\exp\left(  -\frac{i\sigma}{2}\sum_{j=1}^{N}\Delta
\theta_{j}\right)  \left\langle \psi_{\sigma};\phi_{i}|s\right\rangle
_{\vec{r}_{i}}\left\vert \psi_{\sigma};\phi_{d}\right\rangle \text{ ,}%
\end{align}
where we have returned the notations for the RD eigenspinors $\left\vert
\psi_{\sigma};\phi_{i}\right\rangle =\left\vert \psi_{\sigma^{\prime}%
}\right\rangle $ and $\left\vert \psi_{\sigma};\phi_{d}\right\rangle
=\left\vert \psi_{\sigma^{\left(  N\right)  }}\right\rangle $, and also the
input spin $\left\vert s\right\rangle _{\vec{r}_{i}}=\left\vert s\right\rangle
_{\vec{r}_{1}}$. Rewriting the prefactor and the total phase into integration
forms%
\begin{align}
\exp i\left(  \sum_{n=1}^{N}\bar{k}_{n}\Delta r\right)   &  \rightarrow\exp
i\int\limits_{C}\bar{k}\left(  \phi\right)  dr\text{ ,}\\
\sum_{j=1}^{N}\Delta\theta_{j}\left.  =\right.  \sum_{j=1}^{N}\frac
{2m_{j}^{\star}\zeta_{j}}{\hbar^{2}}\Delta r &  \rightarrow\frac{2}{\hbar^{2}%
}\int\limits_{C}m^{\star}\zeta dr\equiv\Delta\Theta\text{ ,}%
\end{align}
we finally obtain%
\begin{align}
\left\vert s\right\rangle _{\vec{r}_{i}\rightarrow\vec{r}_{d}}^{C}  &
=e^{i\int\nolimits_{C}\bar{k}\left(  \phi\right)  dr}\nonumber\\
& \times\sum_{\sigma}\exp\left(  -\frac{i\sigma}{2}\Delta\Theta\right)
\left\langle \psi_{\sigma};\phi_{i}|s\right\rangle _{\vec{r}_{i}}\left\vert
\psi_{\sigma};\phi_{d}\right\rangle \text{ .}\label{sket_nonuni_original}%
\end{align}
Since the prefactor does not survive when doing expectation values (such as
$\left\langle \vec{\sigma}\right\rangle $) with the above ket, it makes no
difference to express $\left\vert s\right\rangle _{\vec{r}_{i}\rightarrow
\vec{r}_{d}}^{C}$ as Eq. (\ref{sket_nonuni}).

Although we have shown that the common phases in Eqs.
(\ref{sket_uniform_original}) and (\ref{sket_nonuni_original}) are
\emph{irrelevant} when calculating, e.g., the spin-vectors $\langle\vec
{S}\rangle$, we remind here that these phases will become \emph{relevant} once
superposition of different states is involved. In that case the prefactor
induces oscillations due to interference between the superposed states, and is
therefore not negligible, unless the inverse Fermi momentum is much larger
than the system size, i.e., $\bar{k}\Delta r\ll1$.

\end{document}